\documentstyle[12pt]{article}

\advance\textheight by \topskip
\newcommand{\sla}{\kern -5.4pt /}
\newcommand{\be}{\begin{equation}}
\newcommand{\ee}{\end{equation}}
\newcommand{\bea}{\begin{eqnarray}}
\newcommand{\eea}{\end{eqnarray}}
\newcommand{\beanon}{\begin{eqnarray*}}
\newcommand{\eeanon}{\end{eqnarray*}}
\newcommand{\ba}{\begin{array}}
\newcommand{\ea}{\end{array}}
\newcommand{\bi}{\begin{itemize}}
\newcommand{\ei}{\end{itemize}}
\newcommand{\ben}{\begin{enumerate}}
\newcommand{\een}{\end{enumerate}}
\newcommand{\bc}{\begin{center}}
\newcommand{\ec}{\end{center}}

\newcommand{\ol}{\overline}
\newcommand{\dotp}{\!\cdot\!}

\newcommand{\NP}[1]{{\it Nucl.\ Phys.\ }{\bf #1}}
\newcommand{\PL}[1]{{\it Phys.\ Lett.\ }{\bf #1}}

\newcommand{\AP}[1]{{\it Ann.\ Phys.\ }{\bf #1}}

\newcommand{\PR}[1]{{\it Phys.\ Rev.\ }{\bf #1}}

\newcommand{\HPA}[1]{{\it Helv.\ Phys.\ Acta.\ }{\bf #1}}

\begin{document}
\tolerance=100000
\input feynman
\thispagestyle{empty}
\setcounter{page}{0}

\begin{flushright}
{\large DFTT 76/93}\\
{\rm January 1994\hspace*{.5 truecm}}\\
\end{flushright}

\vspace*{\fill}

\bc
{\Large \bf A new method for helicity calculations
\footnote{ Work supported in part by Ministero
dell' Universit\`a e della Ricerca Scientifica.\\[2 mm]
e-mail: ballestrero,maina@to.infn.it}}\\[2.cm]
{\large Alessandro Ballestrero and Ezio Maina}\\[.3 cm]
{\it Dipartimento di Fisica Teorica, Universit\`a di Torino, Italy}\\
{\it and INFN, Sezione di Torino, Italy}\\
{\it v. Giuria 1, 10125 Torino, Italy.}\\
\ec

\vspace*{\fill}

\begin{abstract}
{\normalsize
\noindent
We propose a new helicity formalism based on the formal insertion in
spinor lines of a complete set of states build up with unphysical spinors.
The method is developed both for massless and massive fermions for which it
turns  out to be particularly fast. All relevant formulae are given. }
\end{abstract}

\vspace*{\fill}

\newpage
\subsection*{Introduction}
In high energy collisions many particles or partons widely separated
in phase space are often produced. The calculation of  cross
sections for these processes is made difficult by the large number of
Feynman diagrams which appear in the perturbative expansion.
This is due both to the complexity of non abelian theories and to simple
combinatorics,  which  generates more and more diagrams  when  the
number of external particles grows.
\par
If one computes unpolarized cross sections for a process described by many
Feynman diagrams with the textbook method of considering
the amplitude modulus
squared   and taking the traces, one can end up
with a prohibitive number of traces to evaluate.
The calculation becomes  simpler if one uses
the so called helicity-amplitudes techniques. In such an approach, given
an assigned helicity to external particles, one computes the
contribution of every single diagram $k$ as a complex number $a_k$,
sums over all $k$'s and takes the modulus squared. To obtain
unpolarized cross sections one simply sums the modulus squared for
the various external helicities.
\par
The use of helicity amplitude techniques in high energy physics
dates back to ref.~\cite{jw,bj}. Many different approaches have been
developed [3-12],
and even a brief survey of the vast literature on this subject  goes far beyond
the scope of this paper.
\par
Two among the most popular schemes are those of ref.~\cite{ks,hz}.
They can be used both with massless and massive particles.
We have employed them in the past for several calculations.
One of them  is based \cite{hz}  on
the choice of a specific representation for the Dirac matrices. It is then
possible to obtain explicit expressions for the spinors, for the matrices at
the vertices and for the fermion propagators. The complex number corresponding
to a given diagram is obtained multiplying these matrices. In the
other one \cite{ks,mana} all spinors for any physical momentum are defined in
term
of a basic spinor for an  auxiliary lightlike momentum. Decomposing the
internal momenta in terms of the external ones, and using the fact that
$\sum_\lambda u\bar u=p\sla +m$, $\sum_\lambda v\bar v =p\sla-m$,
all spinor lines are reduced to an algebraic combination of spinors products
$\bar u(p_1)u(p_2)$. In order to use this method the polarization vectors of
spin-1 particles have to be expressed as $\bar u\gamma^\mu u$ currents.
\par
In the rest of the paper we describe an approach to helicity amplitudes
which is based on the formal insertion in
spinor lines of a complete set of states build up with unphysical spinors. This
new formalism, that we have tested in several physical computations,
combines the best features of \cite{ks,hz}, is highly flexible,
has a modular structure and in our experience is faster than previous methods.

\subsection*{Spinors}
The method we will describe in the following sections makes use of
generalized  spinors $u(p)$ which coincide with the usual
ones when $p^2\geq 0$, but are defined also for spacelike momenta.
We discuss such a generalization in this section.

Spinors may be defined as eigenstates of $p\sla$. This is equivalent to say
that they must satisfy Dirac equations:
\be\label{dirac}
p\sla u(p)=+m u(p) \hskip 2 truecm
p\sla v(p)=-m v(p)
\ee
Since $p\sla$ is not hermitian, $m$ need not to be real, but it is constrained
to satisfy $m^2=p^2$ as $p\sla^2=p^2$. When $p^2\leq 0$, the two
eigenvalues are imaginary and one can choose, for instance, to  associate
$u(p)$ with the eigenvalue $m$ whose imaginary part is positive.
The remaining degeneracy is normally  eliminated considering eigenstates
of $\gamma^5 s\sla$ :
\bea
\gamma^5 s\sla u(p,+)=u(p,+) \hskip 2 truecm \gamma^5 s\sla u(p,-)=-u(p,-)
\nonumber\\
\gamma^5 s\sla v(p,+)=v(p,+) \hskip 2 truecm \gamma^5 s\sla v(p,-)=-v(p,-)
\label{g5s}
\eea
where the polarization vector $s$ has to satisfy the two conditions
\be\label{s conditions}
s\dotp p=0  \hskip 2 truecm  s^2=-1
\ee
The first one implies that $\gamma^5 s\sla$ commutes with $p\sla$, the
second that $(\gamma^5 s\sla)^2=1$ and hence that its two  eigenvalues
are $\pm 1$.
\par
One can easily costruct an example of spinors satisfying
eqs.(\ref{dirac},\ref{g5s}) for any value of $p^2$
with a straightforward generalization of the method of ref.\cite{ks}. In such a
method, one first defines spinors $w(k_0,\lambda)$ for an auxiliary
massless vector $k_0$ satisfying
\be\label{wspinor}
w(k_0,\lambda)\bar{w}(k_0,\lambda)=\frac{1+\lambda\gamma_5}{2}k\sla_0
\ee
and with their relative phase fixed by
\be
w(k_0,\lambda)=\lambda k\sla_1 w(k_0,-\lambda),
\ee
with $k_1$  a second auxiliary vector such that $k_1^2=-1$, $k_0\dotp k_1=0$.
Spinors for a  four momentum $p$, with
$m^2=p^2$ are then obtained as:
\be\label{uvks}
u(p,\lambda)=\frac{p\sla + m}{\sqrt{2\,p\dotp k_0}}\;w(k_0,-\lambda)
\hskip 1 truecm
v(p,\lambda)=\frac{p\sla - m}{\sqrt{2\,p \dotp k_0}}\;w(k_0,-\lambda)
\ee
One can readily  check that $u$ and $v$ of eq.(\ref{uvks}) satisfy
eqs.(\ref{dirac}, \ref{g5s}) also when $p^2\leq 0$ and $m$ is imaginary.
\par
Some care must be taken in defining in the general case the conjugate spinors
$\bar u(p,\lambda)$, $\bar v(p,\lambda)$. In fact the usual quantities
$\bar u=u^\dagger\gamma^0$, $\bar v=v^\dagger\gamma^0$ do not satisfy normal
Dirac equations
\be\label{diracconj}
\bar u(p) p\sla =+m \bar u(p) \hskip 2 truecm
\bar v(p) p\sla =-m \bar v(p)
\ee
when $m$ is imaginary. Taking the hermitiean conjugate of eqs.(\ref{dirac}),
one readily verifies on the contrary that $\bar u=v^\dagger\gamma_0$,
$\bar v=u^\dagger\gamma_0$ do.
\par
We define $\bar u(p,\lambda)$, $\bar v(p,\lambda)$ as the spinors
 satisfiyng both eqs.(\ref{diracconj}) and
\bea
\bar u(p,+) \gamma^5 s\sla =\bar u(p,+) \hskip 2 truecm
\bar u(p,-) \gamma^5 s\sla =-\bar u(p,-)
\nonumber\\
\bar v(p,+) \gamma^5 s\sla =\bar v(p,+) \hskip 2 truecm
\bar v(p,-) \gamma^5 s\sla =-\bar v(p,-)
\label{g5sconj}
\eea
With this definition the usual  orthogonality relations
among spinors are satisfied. As a consequence, choosing the normalization
\be
\bar u(p,\lambda)u(p,\lambda)=2m \hskip 1 truecm
\bar v(p,\lambda)v(p,\lambda)=-2m,
\ee
the completeness relation
\be\label{completeness}
1=\sum_\lambda\frac{u(p,\lambda)\bar{u}(p,\lambda)-v(p,\lambda)
\bar{v}(p,\lambda)}{2m}
\ee
is verified for any value of $p^2$.
\par
The explicit relation between $u$, $v$ and $\bar u$, $\bar v$ is
\be
\bar u(p,\lambda)=u^\dagger(p,\lambda) \gamma^0\hskip 1 truecm
\bar v(p,\lambda)=v^\dagger(p,\lambda) \gamma^0
\ee
as usual when $p^2>0$. For $p^2<0$ one has
\be
\bar u(p,\lambda)=v^\dagger(p,\lambda) \gamma^0\hskip 1 truecm
\bar v(p,\lambda)=u^\dagger(p,\lambda) \gamma^0
\ee
if the components of the polarization vector $s$ are real, and
\be
\bar u(p,\lambda)=v^\dagger(p,-\lambda) \gamma^0\hskip 1 truecm
\bar v(p,\lambda)=u^\dagger(p,-\lambda) \gamma^0
\ee
if  they become purely imaginary for imaginary $m$.
This is precisely the case when one uses the spinors of ref.\cite{ks}:
the polarization vector is $s^\mu=p^\mu/m-(m/p\dotp k_0)k_0^\mu$
and for a spacelike momentum $p$ it is equal to the imaginary unit times
a timelike vector.
It is interesting to notice that for $k_0^0=\alpha (p^0-|p|)/p^2$ and
$\vec k_0=\alpha \vec p (|p|-p^0)/|p|p^2$, with $\alpha$ an
arbitrary factor, one recovers  the usual helicity polarization vector
$s=(|p|^2,p_0\vec p)/m |p|$.

With the previous definitions, the spinors conjugate to $u$ and $v$
defined in eq.(\ref{uvks}) are given by the simple formulae
\be\label{uvksconj}
\bar u(p,\lambda)=\bar w(k_0,-\lambda)\;\frac{p\sla + m}{\sqrt{2\,p \dotp k_0}}
\hskip 1 truecm
\bar v(p,\lambda)=\bar w(k_0,-\lambda)\;\frac{p\sla - m}{\sqrt{2\,p \dotp k_0}}
\ee
in all cases. Moreover, making use of spinors of such a kind to compute
amplitudes, one has never
in practice  to worry about which determination of $\sqrt{p^2}$ and
$\sqrt{2\,p \dotp k_0}$ to use when $p$ is spacelike.
As we will see in the next sections in fact, only the square of the
above quantities appear at the end of the computation.

\subsection*{Outline of the method. $T$ functions}
The spinor part of every massive fermion line with $n$ insertions

\begin{picture}(34000,8000)
\THICKLINES
\drawline\fermion[\E\REG](3000,2000)[34000]
\drawarrow[\W\ATTIP](\pmidx,\pmidy)
\global\advance \pmidx by -1000
\global\advance \pmidy by -1300
\put (\pmidx,\pmidy){\ldots \ldots}
\global\advance \pmidy by -3000
\put (\pmidx,\pmidy){Fig. 1}
\global\advance \pmidy by 6300
\put (\pmidx,\pmidy){\ldots \ldots}
\global\advance \pfrontx by 2358
\put (\pfrontx,500){$p_1$}
\THINLINES
\global\advance \pfrontx by 2358
\global\seglength=1000 \global\gaplength=250
\drawline\scalar[\N\REG](\pfrontx,\pfronty)[3]
\global\advance \pfrontx by 2858
\put (\pfrontx,500){$p_2$}
\global\seglength=1000 \global\gaplength=250
\global\advance \fermionfrontx by 10432
\drawline\scalar[\N\REG](\fermionfrontx,\fermionfronty)[3]
\global\advance \pfrontx by 2358
\put (\pfrontx,500){$p_3$}
\global\seglength=1000 \global\gaplength=250
\global\advance \fermionbackx by -4716
\drawline\scalar[\N\REG](\fermionbackx,\fermionbacky)[3]
\global\advance \pfrontx by 2358
\put (\pfrontx,500){$p_{n+1}$}
\global\seglength=1000 \global\gaplength=250
\global\advance \fermionbackx by -5716
\drawline\scalar[\N\REG](\fermionbackx,\fermionbacky)[3]
\global\advance \pfrontx by 2358
\put (\pfrontx,500){$p_{n}$}
\end{picture}
\vskip 1.5 truecm

\noindent
has a generic expression of the following type
\be
T^{(n)}=
\ol{U}(p_1,\lambda_1)\chi_1(p\sla_2+\mu_2)\chi_2(p\sla_3+\mu_3)\cdots
(p\sla_{n}+\mu_{n})\chi_{n}U(p_{n+1},\lambda_{n+1})
\ee
where $\lambda_1$ and $\lambda_{n+1}$ are the polarizations of the external
fermions,  $p_1$ and $p_{n+1}$  their momenta. $p_2,\ldots ,p_{n}$ and
$\mu_2,\ldots ,\mu_{n}$ are the 4-momenta and masses appearing in the fermion
propagators. $U(p,\lambda)$ ($\ol{U}(p,\lambda)$) stands for either
 $u(p,\lambda)$ ($\bar{u}(p,\lambda)$)
or  $v(p,\lambda)$ ($\bar{v}(p,\lambda)$).
The $\chi$'s are
\be\label{chis}
\chi_i^S\equiv \chi^S(c_{r_i},c_{l_i})= c_{r_i} \left(\frac{1+\gamma_5}{2}
       \right) +c_{l_i} \left(\frac{1-\gamma_5}{2}\right)
\ee
when the  insertion corresponds to a scalar (or pseudoscalar), or
\be\label{chiv}
\chi_i^V\equiv \chi^V(\eta_i,c_{r_i},c_{l_i})= \eta\sla_i \left[
c_{r_i} \left(\frac{1+\gamma_5}{2}\right) +
             c_{l_i} \left(\frac{1-\gamma_5}{2}\right) \right]
\ee
when it corresponds to a vector particle whose `polarization' is $\eta$.
Of course $\eta$ can be the polarization vector of the external particle
or the vector resulting from a complete subdiagram which is connected
in the {\it i}--th position to the fermion line.

Let us start considering the case in which there are only two insertions:
\be\label{T2}
T^{(2)}(p_1;\eta_1,c_1;p_2;\eta_2,c_2;p_3)=\ol{U}(p_1,\lambda_1)\chi_1(p\sla_2+
\mu_2)\chi_2 U(p_3,\lambda_3).
\ee
Here and in the following we indicate with $c_i$ both the couplings $c_{r_i}$
and $c_{l_i}$. Obviously the vectors $\eta$ only appear as an argument for
vector insertions.
\par
Even if $m_2^2\equiv p_2^2$ does not correspond to the mass of any physical
particle, one can insert in eq. (\ref{T2}), just before $(p\sla_2+\mu_2)$
a complete set  of states in the form:
\be
1=\sum_\lambda\frac{u(p_2,\lambda)\bar{u}(p_2,\lambda)-v(p_2,\lambda)
\bar{v}(p_2,\lambda)}{2m_2}
\ee
and make use of Dirac equations to get:
\bea
T^{(2)} & = &\frac{1}{2}\sum_{\lambda_2}\left(\ol{U}(p_1,\lambda_1)\chi_1
u(p_2,\lambda_2)\times
\bar{u}(p_2,\lambda_2)\chi_2 U(p_3,\lambda_3)\times \left( 1+{\mu_2\over
m_2}\right)\right.\nonumber \\ \label{T2uv}
    &   &
+\left.\ol{U}(p_1,\lambda_1)\chi_1v(p_2,\lambda_2)\times
\bar{v}(p_2,\lambda_2)\chi_2 U(p_3,\lambda_3)\times \left( 1-{\mu_2\over
m_2}\right)\right)
\eea
This example can  be generalized to any number of insertions and shows
that the factors $(p\sla_i+\mu_i)$ can be easily eliminated,
reducing all fermion lines essentially to  products of $T$ functions:
\be\label{T}
T_{\lambda_1 \lambda_2}(p_1;\eta,c;p_2)=\ol{U}(p_1,\lambda_1)\chi U(p_2,
\lambda_2)
\ee
defined for any value of $p_1^2$ and $p_2^2$.

It is convenient to use the spinors  $u(p,\lambda)$ and $v(p,\lambda)$
defined in eqs.(\ref{uvks},\ref{uvksconj}). With this choice, to which we will
adhere
from now on, the $T$ functions (\ref{T}) have a simple dependence on $m_1$
and $m_2$ and as a consequence the rules for constructing  spinor lines out
of them are simple.
Every  $T$ function has in fact an expression of the following kind:
\be\label{Texpr}
\widetilde T_{\lambda_1 \lambda_2}(p_1;\eta,c;p_2)\equiv
\sqrt{p_1\dotp k_0}\;\sqrt{p_2\dotp k_0}\;
T_{\lambda_1 \lambda_2}(p_1;\eta,c;p_2)=\hskip 4truecm
\ee
\[
A_{\lambda_1 \lambda_2}(p_1;\eta,c;p_2)
+M_1B_{\lambda_1 \lambda_2}(p_1;\eta,c;p_2)
 +M_2C_{\lambda_1 \lambda_2}(p_1;\eta,c;p_2)
+M_1M_2D_{\lambda_1 \lambda_2}(p_1;\eta,c;p_2)\nonumber
\]
where
\bea\label{UM}
M_i=+m_i \hskip 1truecm \mbox{if}\hskip 1truecm
U(p_i,\lambda_i)=u(p_i,\lambda_i) \\
M_i=-m_i \hskip 1truecm \mbox
{if}\hskip 1truecm
U(p_i,\lambda_i)=v(p_i,\lambda_i).\nonumber
\eea
The functions $A$, $B$, $C$, $D$ turn out to be independent of
$m_1$ and $m_2$ and of the $u$ or $v$ nature of $\ol{U}(p_1,\lambda_1)$
and $U(p_2,\lambda_2)$.
We give in Appendix A the  expressions for $A^V$, $B^V$, $C^V$,
$D^V$ and $A^S$, $B^S$, $C^S$, $D^S$, which are the $A$, $B$, $C$, $D$
functions for a vector and a scalar insertion respectively .

\subsection*{Spinor lines}
We will show in this section how to compute recursively the functions
\be\label{titilde}
\widetilde T^{(n)}=T^{(n)} \sqrt{p_1\dotp k_0}\;\sqrt{p_{n+1}\dotp k_0}
\; (p_2\dotp k_0) \; (p_3\dotp k_0) \cdots \; (p_n\dotp k_0).
\ee
from which the $T^{(n)}$ themselves can then be immediately obtained
at the end of the computation, dividing by the appropriate factors.
\par
Let us denote with $\widetilde T,A,B,C,D$ the $2x2$ matrices whose elements are
$\widetilde T_{\lambda_1 \lambda_2}$, $A_{\lambda_1 \lambda_2}$,
$B_{\lambda_1\lambda_2}$,
$C_{\lambda_1 \lambda_2}$, $D_{\lambda_1 \lambda_2}$.
With this notation, making use of eqs. (\ref{Texpr}) and (\ref{T}),
eq. (\ref{T2uv}) reads:
\bea
\lefteqn{ \widetilde T^{(2)}(1,2,3)  =
\frac{1}{2}\biggl[\Bigl(A(1,2)+M_1B(1,2)+m_2C(1,2)+
M_1m_2D(1,2)\Bigr)}\nonumber \\
& & \times \left( 1+{\mu_2\over m_2}\right) \times
 \Bigl(A(2,3)+m_2B(2,3)+M_3C(2,3)+m_2M_3D(2,3)\Bigr)\label{multi}\\
& & + \Bigl(A(1,2)+M_1B(1,2)-m_2C(1,2)-M_1m_2D(1,2)\Bigr)
  \nonumber \\
& & \times \left( 1-{\mu_2\over m_2}\right)\times
  \Bigl(A(2,3)-m_2B(2,3)+M_3C(2,3)-m_2M_3D(2,3)\Bigr)\biggr]
\nonumber
\eea
where we have used the shorthands $(1,2)$ and $(1,2,3)$ for
$(p_1;\eta_1,c_1;p_2)$ and $(p_1;\eta_1,c_1;p_2;\eta_2,c_2;p_3)$ respectively.
\par
Elementary algebra shows that $\widetilde T^{(2)}$ has again the same
dependence on the
external (possibly unphysical) masses as in (\ref{Texpr}):
\be\label{T2/2}
\widetilde T^{(2)}(1,2,3)=A^{(2)}(1,2,3)+M_1B^{(2)}(1,2,3)+M_3C^{(2)}(1,2,3)
+M_1M_3D^{(2)}(1,2,3)
\ee
with
\bea
A^{(2)}(1,2,3)=A(1,2)\Bigl(A(2,3)+\mu_2B(2,3)\Bigr)
  +C(1,2)\Bigl(\mu_2A(2,3)+p_2^2B(2,3)\Bigr)\nonumber\\
B^{(2)}(1,2,3)=B(1,2)\Bigl(A(2,3)+\mu_2B(2,3)\Bigr)
  +D(1,2)\Bigl(\mu_2A(2,3)+p_2^2B(2,3)\Bigr)\label{abcd}\\
C^{(2)}(1,2,3)=A(1,2)\Bigl(C(2,3)+\mu_2D(2,3)\Bigr)
  +C(1,2)\Bigl(\mu_2C(2,3)+p_2^2D(2,3)\Bigr)\nonumber\\
D^{(2)}(1,2,3)=B(1,2)\Bigl(C(2,3)+\mu_2D(2,3)\Bigr)
  +D(1,2)\Bigl(\mu_2C(2,3)+p_2^2D(2,3)\Bigr)\nonumber
\eea
This implies that $A^{(2)}$, $B^{(2)}$, $C^{(2)}$, $D^{(2)}$ can be reinserted
in an equation like eq. (\ref{multi}) to give the $\widetilde T$
function $\widetilde T^{(3)}$
corresponding to a fermion line with 3 insertions, and so on.
So one can generalize
eqs. (\ref{multi},\ref{T2/2},\ref{abcd})  by induction: every
$\widetilde T^{(i)}$ turns out to be of the form
\be\label{Ti}
\widetilde T^{(i)}=A^{(i)}+M_1B^{(i)}+M_{i+1}C^{(i)}+M_1M_{i+1}D^{(i)}
\ee
and from the knowledge of $A^{(i)}(1,\ldots,i)$, $B^{(i)}(1,\ldots,i)$,
$C^{(i)}(1,\ldots,i)$, $D^{(i)}(1,\ldots,i)$ and $A^{(j)}(i+1,\ldots,i+j)$,
$B^{(j)}(i+1,\ldots,i+j)$, $C^{(j)}(i+1,\ldots,i+j)$,
$D^{(j)}(i+1,\ldots,i+j)$,  one gets:
\bea
A^{(i+j)}=A^{(i)}\Bigl(A^{(j)}+\mu_iB^{(j)}\Bigr)
  +C^{(i)}\Bigl(\mu_iA^{(j)}+p_i^2B^{(j)}\Bigr)\nonumber\\
B^{(i+j)}=B^{(i)}\Bigl(A^{(j)}+\mu_iB^{(j)}\Bigr)
  +D^{(i)}\Bigl(\mu_iA^{(j)}+p_i^2B^{(j)}\Bigr)\label{abcd/2}\\
C^{(i+j)}=A^{(i)}\Bigl(C^{(j)}+\mu_iD^{(j)}\Bigr)
  +C^{(i)}\Bigl(\mu_iC^{(j)}+p_i^2D^{(j)}\Bigr)\nonumber\\
D^{(i+j)}=B^{(i)}\Bigl(C^{(j)}+\mu_iD^{(j)}\Bigr)
  +D^{(i)}\Bigl(\mu_iC^{(j)}+p_i^2D^{(j)}\Bigr)\nonumber
\eea
which build up $\widetilde T^{(i+j)}(1,\ldots,i+j)$. In eq.~(\ref{abcd/2})
$\mu_i$ and $p_i$ are the mass and
momentum of the propagator which connects the left $i$ insertions with the
right $j$ ones
\par
It should by now be evident that the evaluation of any spinor line can be
performed
computing the $A$, $B$, $C$, $D$ matrices relative to every single insertion
and combining them toghether with the help of eq. (\ref{abcd/2}) until one gets
to the final $T^{(n)}$.
It is important to point out that the unphysical masses $m_i$
($i=2,\cdots,n$) do not appear
in eqs. (\ref{abcd},\ref{Ti},\ref{abcd/2}): only their squares $p_i^2$ do.
Therefore,as anticipated, the determination of $m_i=\sqrt{p_i^2}$ is
irrelevant.
The same conclusion can be drawn for $\sqrt{p_{i}\dotp k_0}$
 from
eq.~(\ref{titilde}) and the fact that the expressions~(\ref{a2})
are independent of square roots.

\subsection*{$\tau$ matrices}
It is convenient to cast the previous formulae in a matrix notation. We drop
the superscripts $(n)$ when not necessary.
\par
Every piece of a spinor line as well as every complete spinor line with $n$
insertions is completely known when we know the matrix
\be
\tau=\left(\ba{cc}A&C\\B&D\ea\right)
\ee
The law of composition  of two pieces of spinor line ,
connected by a fermion propagator with 4-momentum $p$ and mass $\mu$,
whose matrices are
\[
\tau_1=\left(\ba{cc}A_1&C_1\\B_1&D_1\ea\right)
\hskip 2truecm
\tau_2=\left(\ba{cc}A_2&C_2\\B_2&D_2\ea\right) ,
\]
is simply (cfr. eq. (\ref{abcd/2}) ) :
\be\label{comp}
\left(\ba{cc}A&C\\B&D\ea\right)  =
\left(\ba{cc}A_1&C_1\\B_1&D_1\ea\right)
\left(\ba{cc}1&\mu\\\mu&p^2\ea\right)
\left(\ba{cc}A_2&C_2\\B_2&D_2\ea\right)
\ee
We will sometimes indicate the above composition as follows:
\be
\tau=\tau_1\bullet\tau_2
\ee
If we call $\pi_i$ the matrix
\be
\pi_i=\left(\ba{cc}1&\mu_i\\\mu_i&p_i^2\ea\right)
\ee
corresponding to the propagator of 4-momentum $p_i$
and $\tau_i$ the matrix associated with the $i$-th insertion of fig.~1,
the $\tau$ matrix  of the whole spinor line can  then be computed as follows:
\be\label{tau}
\tau=\tau_1\pi_2\tau_2\pi_3\tau_3\cdots\pi_{n-1}\tau_{n-1}\pi_{n}\tau_{n}
\ee
Explicitely written as  $4\times 4$ matrices,  $\tau$ and $\pi_i$ are:
\be\label{taupi}
\tau=\left(\ba{cccc}A_{++}&A_{+-}&C_{++}&C_{+-}\\
                    A_{-+}&A_{--}&C_{-+}&C_{--}\\
                    B_{++}&B_{+-}&D_{++}&D_{+-}\\
                    B_{-+}&B_{--}&D_{-+}&D_{--}\ea\right)
\hskip 1 truecm
\pi_i=\left(\ba{cccc}  1  &  0  &\mu_i&  0  \\
                       0  &  1  &  0  &\mu_i\\
                     \mu_i&  0  &p_i^2&  0  \\
                       0  &\mu_i&   0   &p_i^2\ea\right)
\ee
 From
the espressions of $A$, $B$, $C$, $D$ given in appendix A, one can see
that the $\tau$'s of a single vector or scalar insertion have the
particular form:
\be\label{tauvs}
\tau^V=\left(\ba{cccc}A_{++}^V&   0    &   0    &C_{+-}^V\\
                         0    &A_{--}^V&C_{-+}^V&   0    \\
                         0    &B_{+-}^V&D_{++}^V&   0    \\
                      B_{-+}^V&   0    &   0    &D_{--}^V\ea\right)
\hskip 1truecm
\tau^S=\left(\ba{cccc}   0    &A_{+-}^S&C_{++}^S&   0    \\
                      A_{-+}^S&   0    &   0    &C_{--}^S\\
                      B_{++}^S&   0    &   0    &   0    \\
                         0    &B_{--}^S&   0    &   0    \ea\right)
\ee
When the insertion correspond to a  $W$ boson, only $A_{--}^V$, $C_{-+}^V$,
$B_{+-}^V$, $D_{++}^V$ are different from zero. For practical computations we
have implemented a set of routines which automatically write lines of Fortran
code both for the expressions of the $A$, $B$, $C$, $D$ functions
and for their combination to form whole spinor lines. These routines of course
avoid unnecessary and time consuming multiplications by the zeroes of the
$\tau$ matrices.
\par
Before ending the section we just notice that the $4\times 4$ matrices $\tau$
and $\pi$ could also be defined exchanging indices $2$ and $3$, so that
eqs.~(\ref{taupi}) become:
\be\label{taupinew}
\tau=\left(\ba{cc}\tau_{++}&\tau_{+-}\\
                  \tau_{-+}&\tau_{--}\ea\right)
    =\left(\ba{cccc}A_{++}&C_{++}&A_{+-}&C_{+-}\\
                    B_{++}&D_{++}&B_{+-}&C_{+-}\\
                    A_{-+}&C_{-+}&A_{--}&C_{--}\\
                    B_{-+}&D_{-+}&B_{--}&D_{--}\ea\right)
\ee

\be
\pi_i=\left(\ba{cccc}  1  &\mu_i&  0  &  0  \\
                     \mu_i&  1  &  0  &  0  \\
                       0  &  0  &p_i^2&\mu_i\\
                       0  &  0  &\mu_i&p_i^2\ea\right)
\ee

\subsection*{Massless spinor lines}
When one has to deal with massless spinor lines, all the formulae given in the
appendix A remain valid. But in this case the fact that all $\mu_i$'s as
well as $m_1$ and $m_{n+1}$ are zero leads to significant simplifications.
For example
the rule (\ref{abcd/2}) for combining pieces of spinor line is now
\bea
A^{(i+j)}=A^{(i)} A^{(j)} +p_i^2 C^{(i)} B^{(j)}\nonumber\\
B^{(i+j)}=B^{(i)} A^{(j)} +p_i^2 D^{(i)} B^{(j)}\label{abcd0}\\
C^{(i+j)}=A^{(i)} C^{(j)} +p_i^2 C^{(i)} D^{(j)}\nonumber\\
D^{(i+j)}=B^{(i)} C^{(j)} +p_i^2 D^{(i)} D^{(j)}\nonumber
\eea
and the
$\pi_i$ matrices (\ref {tau}, \ref {taupi}) become diagonal:
\be
\pi_i=
\left(\ba{cccc}  1  &  0  &0&  0  \\
                       0  &  1  &  0  &0\\
                     0&  0  &p_i^2&  0  \\
                       0  &0&   0   &p_i^2\ea\right)
\ee

Moreover, in order to compute the whole spinor line with $n$ insertions
it is not necessary to know
the whole $\tau^{(n)}$ matrix. It is clear from eq.~(\ref{Ti}) that only
$A^{(n)}$ is needed. Therefore, if one for instance  multiplies recursively
the $\tau$ matrices of the single insertions starting from the left,
at every single step one needs to compute only $A$ and $C$.
In fact, from eq.~(\ref{abcd0}) with $i=n-1$, $j=1$, one sees that
$B^{(n-1)}$ and $D^{(n-1)}$ are not needed to compute $A^{n}$. With
$i=n-2$, $j=1$ one verifies that $B^{(n-2)}$ and $D^{(n-2)}$ are not
needed to compute $A^{(n-1)}$ and $C^{(n-1)}$, and so on.
Had one started
from the right, only $A$ and $B$ would have had  to be calculated for
every product.

In most theories, like e.g. the standard model, one has to consider only
vector and axial-vector couplings to  massless spinor lines. This means
that one has to compute only $\tau^V$ matrices (\ref{tauvs}) for every
insertion. Combining together two matrices of this type, one still gets a
matrix whose only elements different from zero are on the two
diagonals:

\beanon
\left(\ba{cccc}       A_{++}  &   0  &   0  &C_{+-}\\
                         0    &A_{--}&C_{-+}&  0   \\
                         0    &B_{+-}&D_{++}&  0   \\
                      B_{-+}&   0    &   0    &D_{--}\ea\right)
\left(\ba{cccc}  1  &  0  &0&  0  \\
                       0  &  1  &  0  &0\\
                     0&  0  &p_i^2&  0  \\
                       0  &0&   0   &p_i^2\ea\right)
\left(\ba{cccc}A^{\prime}_{++}&   0    &   0    &C^{\prime}_{+-}\\
                         0    &A^{\prime}_{--}&C^{\prime}_{-+}&   0    \\
                         0    &B^{\prime}_{+-}&D^{\prime}_{++}&   0    \\
                 B^{\prime}_{-+}&   0    &   0    &D^{\prime}_{--}\ea\right)=\\
=\left(\ba{cccc}A^{\prime\prime}_{++}&   0    &   0    &C^{\prime\prime}_{+-}\\
                 0    &A^{\prime\prime}_{--}&C^{\prime\prime}_{-+}&   0    \\
                  0    &B^{\prime\prime}_{+-}&D^{\prime\prime}_{++}&   0    \\
        B^{\prime\prime}_{-+}&   0    &   0    &D^{\prime\prime}_{--}\ea\right)
\eeanon

\noindent
and therefore  the $\tau$ matrix of any piece of massless spinor line is
`cross-diagonal' for these theories.

\subsection*{ Combining spinor lines. A simple example}
In the following, whenever in the argument of a $\tau$ matrix or of one of its
elements there will be an index $\mu$ in place of a `polarization vector'
$\eta$, we imply that the components $\eta^\nu$ have been taken to be
$\eta^\nu=g^{\mu\nu}$. In fact every $\tau$ matrix satisfies the relation
\be
\tau (p_1;\eta_1,\cdots;\eta_i,\cdots,p_n)=\eta_{i\mu}\;
\tau (p_1;\eta_1,\cdots;\mu,\cdots,p_n)
\ee
and of course similar relations can be written for its $A$,$B$,$C$,$D$
components. In case of just one insertion we have for example:
\be
A^V_{++}(p_1;\eta,c;p_2)=\eta_\mu\; A^V_{++}(p_1;\mu,c;p_2)
\ee
{}From eq.~(\ref{a2}) it is immediate to see that
\be
A^V_{++}(p_1;\mu,c;p_2)=c_r\left(-p_1\dotp p_2\;k_0^\mu+k_0\dotp p_1\;
p_2^\mu+
    k_0\dotp p_2\; p_1^\mu-
i \epsilon_{\mu\nu\rho\sigma}\, k_0^\nu\, p_1^\rho\, p_2^\sigma \right)\\
\ee
with $\epsilon_{0123}=1$.
\par
As a simple example of how to use in practice the
method just exposed, let us consider the process $e^+e^-\rightarrow t \bar t
g$.
It is described by the diagrams in fig.~2.
\par
\begin{picture}(40000,28000)
\put (19500,0) {Fig. 2}
\THICKLINES
\bigphotons
\global\Xone=300
\global\Xtwo=-900
\global\Xthree=900
\global\Xfour=-2000
\global\Yone=1700
\global\Ytwo=-1200
\global\Ythree=-1900
\global\Yfour=-1600

\drawline\fermion[\NE\REG](5000,20000)[4000]
\drawarrow[\NE\ATTIP](\pmidx,\pmidy)
\global\advance \pmidy by \Xtwo
\put (\pmidx,\pmidy) {$e^-$}
\drawline\fermion[\NW\REG](\pbackx,\pbacky)[4000]
\drawarrow[\NW\ATTIP](\pmidx,\pmidy)
\global\advance \pmidy by \Xone
\put (\pmidx,\pmidy) {$e^+$}
\drawline\photon[\E\REG](\pfrontx,\pfronty)[5]
\global\advance \pmidx by -600
\global\advance \pmidy by \Ytwo
\put (\pmidx,\pmidy) {$\gamma$}
\drawline\fermion[\NE\REG](\pbackx,\pbacky)[4000]
\drawarrow[\NE\ATTIP](\pmidx,\pmidy)
\global\advance \pmidy by \Xthree
\put (\pmidx,\pmidy) {$t$}
\drawline\fermion[\S\REG](\pfrontx,\pfronty)[4000]
\drawarrow[\N\ATTIP](\pmidx,\pmidy)
\drawline\gluon[\E\FLIPPEDFLAT](\pbackx,\pbacky)[3]
\global\advance \pmidy by \Yone
\put (\pmidx,\pmidy) {$g$}
\drawline\fermion[\SE\REG](\pfrontx,\pfronty)[4000]
\drawarrow[\NW\ATTIP](\pmidx,\pmidy)
\global\advance \pmidy by \Xfour
\put (\pmidx,\pmidy) {$\bar t$}

\drawline\fermion[\NE\REG](26500,20000)[4000]
\drawarrow[\NE\ATTIP](\pmidx,\pmidy)
\global\advance \pmidy by \Xtwo
\put (\pmidx,\pmidy) {$e^-$}
\drawline\fermion[\NW\REG](\pbackx,\pbacky)[4000]
\drawarrow[\NW\ATTIP](\pmidx,\pmidy)
\global\advance \pmidy by \Xone
\put (\pmidx,\pmidy) {$e^+$}
\drawline\photon[\E\REG](\pfrontx,\pfronty)[5]
\global\advance \pmidx by -600
\global\advance \pmidy by \Yfour
\put (\pmidx,\pmidy) {$Z$}
\drawline\fermion[\NE\REG](\pbackx,\pbacky)[4000]
\drawarrow[\NE\ATTIP](\pmidx,\pmidy)
\global\advance \pmidy by \Xthree
\put (\pmidx,\pmidy) {$t$}
\drawline\fermion[\S\REG](\pfrontx,\pfronty)[4000]
\drawarrow[\N\ATTIP](\pmidx,\pmidy)
\drawline\gluon[\E\FLIPPEDFLAT](\pbackx,\pbacky)[3]
\global\advance \pmidy by \Yone
\put (\pmidx,\pmidy) {$g$}
\drawline\fermion[\SE\REG](\pfrontx,\pfronty)[4000]
\drawarrow[\NW\ATTIP](\pmidx,\pmidy)
\global\advance \pmidy by \Xfour
\put (\pmidx,\pmidy) {$\bar t$}

\drawline\fermion[\NE\REG](5000,3000)[4000]
\drawarrow[\NE\ATTIP](\pmidx,\pmidy)
\global\advance \pmidy by \Xtwo
\put (\pmidx,\pmidy) {$e^-$}
\drawline\fermion[\NW\REG](\pbackx,\pbacky)[4000]
\drawarrow[\NW\ATTIP](\pmidx,\pmidy)
\global\advance \pmidy by \Xone
\put (\pmidx,\pmidy) {$e^+$}
\drawline\photon[\E\REG](\pfrontx,\pfronty)[5]
\global\advance \pmidx by -600
\global\advance \pmidy by \Ytwo
\put (\pmidx,\pmidy) {$\gamma$}
\drawline\fermion[\SE\REG](\pbackx,\pbacky)[4000]
\drawarrow[\NW\ATTIP](\pmidx,\pmidy)
\global\advance \pmidy by \Xfour
\put (\pmidx,\pmidy) {$\bar t$}
\drawline\fermion[\N\REG](\pfrontx,\pfronty)[4000]
\drawarrow[\N\ATTIP](\pmidx,\pmidy)
\drawline\gluon[\E\FLAT](\pbackx,\pbacky)[3]
\global\advance \pmidy by \Ythree
\put (\pmidx,\pmidy) {$g$}
\drawline\fermion[\NE\REG](\pfrontx,\pfronty)[4000]
\drawarrow[\NE\ATTIP](\pmidx,\pmidy)
\global\advance \pmidy by \Xthree
\put (\pmidx,\pmidy) {$t$}

\drawline\fermion[\NE\REG](26500,3000)[4000]
\drawarrow[\NE\ATTIP](\pmidx,\pmidy)
\global\advance \pmidy by \Xtwo
\put (\pmidx,\pmidy) {$e^-$}
\drawline\fermion[\NW\REG](\pbackx,\pbacky)[4000]
\drawarrow[\NW\ATTIP](\pmidx,\pmidy)
\global\advance \pmidy by \Xone
\put (\pmidx,\pmidy) {$e^+$}
\drawline\photon[\E\REG](\pfrontx,\pfronty)[5]
\global\advance \pmidx by -600
\global\advance \pmidy by \Yfour
\put (\pmidx,\pmidy) {$Z$}
\drawline\fermion[\SE\REG](\pbackx,\pbacky)[4000]
\drawarrow[\NW\ATTIP](\pmidx,\pmidy)
\global\advance \pmidy by \Xfour
\put (\pmidx,\pmidy) {$\bar t$}
\drawline\fermion[\N\REG](\pfrontx,\pfronty)[4000]
\drawarrow[\N\ATTIP](\pmidx,\pmidy)
\drawline\gluon[\E\FLAT](\pbackx,\pbacky)[3]
\global\advance \pmidy by \Ythree
\put (\pmidx,\pmidy) {$g$}
\drawline\fermion[\NE\REG](\pfrontx,\pfronty)[4000]
\drawarrow[\NE\ATTIP](\pmidx,\pmidy)
\global\advance \pmidy by \Xthree
\put (\pmidx,\pmidy) {$t$}
\end{picture}
\vskip 1. truecm
\noindent
In this case all insertions on spinor lines correspond to vector particles.
We will indicate the four momenta of the particles with
their names, so that
$s=(e^++e^-)^2$. We will also denote with $c^{fV}$ the couplings of a
generic fermion $f$ to  a vector particle $V$.
\par\noindent
One has to choose $k_0$ and $k_1$
 and the form of the polarization vectors $\eta ^g_i$
for the gluon in terms of his momentum. One can for instance use
$k_0=(1,1,0,0)$, $k_1=(0,0,1,0)$ and the real polarization
vectors proposed in ref.~\cite{hz} or normal helicity eigenvectors.
One then starts computing with eqs.~(\ref{a2})
$A_{jj}(e^+;\mu,c^{e\gamma};e^-)$ and $A_{jj}(e^+;\mu,c^{eZ};e^-)$, with
$j(=+,-)$ the polarization of the electron. With them one constructs the
`polarization' vectors
\be
\eta_{j\mu}^{\gamma}=\frac{A_{jj}(e^+;\mu,c^{e\gamma};e^-)}{s} \hskip 2truecm
\eta_{j\mu}^{Z}=\frac{A_{jj}(e^+;\mu,c^{eZ};e^-)}{s-m^2_Z+i\Gamma_Z m_Z}
\ee
Using again eqs.~(\ref{a2}), it is easy to compute the tau matrices
relative to the insertions in the upper and lower part of the line:
\bea
\tau^u_{\gamma} [j]=\tau(t;\eta_j^{\gamma},c^{t\gamma};-\bar t -g)\nonumber\\
\tau^u_{Z} [j]=\tau(t;\eta_j^{Z},c^{tZ};-\bar t -g)\nonumber\\
\tau^u_{g} [i]=\tau(t;\eta_i^{g},c^{tg}; t +g)\nonumber\\
\tau^d_{\gamma} [j]=\tau(t+g;\eta_j^{\gamma},c^{t\gamma};\bar t)\\
\tau^d_{Z} [j]=\tau(t+g;\eta_j^{Z},c^{tZ};\bar t)\nonumber\\
\tau^d_{g} [i]=\tau(-\bar t -g;\eta_i^{g},c^{tg};\bar t)\nonumber
\eea
and then the sum of $\tau_{\gamma}$ and $\tau_{Z}$ matrices:
\be
\tau^u_{\gamma Z} [j]=\tau^u_{\gamma} [j]+\tau^u_{Z} [j]
\hskip 1truecm
\tau^d_{\gamma Z} [j]=\tau^d_{\gamma} [j]+\tau^d_{Z} [j]
\ee
The final $\tau$ matrices are obtained just composing with the law
(\ref{abcd},\ref{comp}) $\tau^u_{\gamma Z} [j]$ with $\tau^d_{g} [i]$ and
$\tau^u_{g} [i]$ with $\tau^d_{\gamma Z} [j]$ :
\be
\tau_1 [i,j]=\tau^u_{\gamma Z} [j] \bullet \tau^d_{g} [i] \hskip 1truecm
\tau_2 [i,j]=\tau^u_{g} [i] \bullet \tau^d_{\gamma Z} [j]
\ee

{}From $\tau [i,j]$ the polarized amplitude is then obtained with the help of
eq.~(\ref{Ti}).
Indicating with $l$ the polarization of the top and with $m$ that of the
antitop, one  has
\bea
\widetilde T_1[i,j,l,m]=A_{1lm} [i,j]+m_t B_{1lm} [i,j]-m_t C_{1lm} [i,j] -
m_t^2 D_{1lm} [i,j]\nonumber\\
\widetilde T_2[i,j,l,m]=A_{2lm} [i,j]+m_t B_{2lm} [i,j]-m_t C_{2lm} [i,j] -
m_t^2 D_{2lm} [i,j]
\eea
and the amplitude
\bea
\mbox {Amp } [i,j,l,m]&=&\frac{1}{\sqrt{e^+\dotp k_0}\;\sqrt{e^-\dotp k_0}\;
\sqrt{t\dotp k_0}\;\sqrt{\bar t\dotp k_0}} \\
&& \hskip-4cm \times \left(\frac
{\widetilde T_1[i,j,l,m]}
{(-\bar t -g)\dotp k_0 \;
((-\bar t -g)^2-m_t^2+i\Gamma_t m_t)} +\frac{\widetilde T_2[i,j,l,m]}
{(t+g)\dotp k_0 \;
((t+g)^2-m_t^2+i\Gamma_t m_t)}\right)\nonumber
\eea
If it were necessary to keep into account the electron mass, one
would have to compute
all the quantities $A_{ij}(e^+;\mu,c^{e\gamma};e^-)$,
$B_{ij}(e^+;\mu,c^{e\gamma};e^-)$,
$C_{ij}(e^+;\mu,c^{e\gamma};e^-)$, $D_{ij}(e^+;\mu,c^{e\gamma};e^-)$ and those
with $c^{eZ}$ replacing $c^{e\gamma}$. Of course
$i$ and $j$, the positron and electron polarization, can be different in this
case. From these one then obtains
\bea
\eta^{\prime\,\gamma}_{ij\mu}&=&A_{ij}(e^+;\mu,c^{e\gamma};e^-)
-m_e B_{ij}(e^+;\mu,c^{e\gamma};e^-)\nonumber\\
&&+m_e C_{ij}(e^+;\mu,c^{e\gamma};e^-)
-m_e^2 D_{ij}(e^+;\mu,c^{e\gamma};e^-)
\eea
\bea
\eta^{\prime\, Z}_{ij\mu}&=&A_{ij}(e^+;\mu,c^{eZ};e^-)
-m_e B_{ij}(e^+;\mu,c^{eZ};e^-)\nonumber\\
&&+m_e C_{ij}(e^+;\mu,c^{eZ};e^-)
-m_e^2 D_{ij}(e^+;\mu,c^{eZ};e^-)
\eea
and, in the unitary gauge with $k=e^++e^-$,
\be
\eta_{ij\mu}^{\gamma}=\frac{\eta^{\prime\, \gamma}_{ij\mu}}{s} \hskip 2truecm
\eta_{ij\mu}^{Z}=\frac{\eta^{\prime\, Z}_{ij\mu}-\eta^{\prime\, Z}_{ij}\dotp k
\,k_\mu/m_Z^2}{s-m^2_Z+i\Gamma_Z m_Z}.
\ee
The rest of the computation is performed in analogy with the massless electron
case.
\par
In general, when one has two spinor lines connected by a vector boson, instead
of computing the `polarization' $\eta_\mu$ of one of the lines and use it in
the
other, one could compute directly the quantity
$\widetilde T_{\lambda_1 \lambda_2}(p_1;\mu,c;p_2)\;
\widetilde T_{\lambda_3 \lambda_4}(p_3;\mu,c^\prime ;p_4)$ where
$p_1$, $p_2$ ($p_3$, $p_4$)
are the momenta of the first (second) spinor line contiguous to the insertion.
This quantity is the analog of the Z functions of ref.~\cite{ks,mana}:
$ Z_{\lambda_1 \lambda_2\lambda_3 \lambda_4} =
T_{\lambda_1 \lambda_2}(p_1;\mu,c;p_2)\;
T_{\lambda_3 \lambda_4}(p_3;\mu,c^\prime ;p_4)$, and it
is linear in all four
(possibly unphysical) masses. In the present formalism one can immediately
derive the formulae for the coefficients of the products of masses just
contracting
$A_{\lambda_1 \lambda_2}(p_1;\mu,c;p_2)$,
$B_{\lambda_1 \lambda_2}(p_1;\mu,c;p_2)$,
$C_{\lambda_1 \lambda_2}(p_1;\mu,c;p_2)$,
$D_{\lambda_1 \lambda_2}(p_1;\mu,c;p_2)$,
with the corresponding
$A_{\lambda_3 \lambda_4}(p_3;\mu,c^\prime ;p_4)$,
$B_{\lambda_3 \lambda_4}(p_3;\mu,c^\prime ;p_4)$,
$C_{\lambda_3 \lambda_4}(p_3;\mu,c^\prime ;p_4)$,
$D_{\lambda_3 \lambda_4}(p_3;\mu,c^\prime ;p_4)$.
Using some algebra to reexpress the determinants (\ref{a3}) which appear
after the contraction, the formulae coincide with the ones one can easily
deduce from the expressions of the Z functions which can be found in the
literature \cite{mana,noi}.
This way of combining the spinor lines is not, in
our opinion, the most convenient one when both spinor lines have several
insertions and when the vector line connecting them has itself some insertions,
due for istance to some triple vector coupling.

\subsection*{Comments and conclusions}
The method we have exposed has been
tested in several computations of physical amplitudes, both unpolarized and
partially or fully polarized. Our results have always been in perfect agreement
with those obtained in other formalisms.
\par
Even if we use for external fermions the same spinors as in
\cite{ks}, we do not have to use also the polarizations they suggest
for vector particles. In effect, expecially for massive vector particles,
we often use the real polarization vectors suggested in \cite{hz}.
If one defines $k_0=(1,0,0,-1), k_1=(0,1,0,0)$, the massless spinors
of  ref.~\cite{ks}
coincide with those of ref.~\cite{hz}.
This implies that with this choice of $k_0$ and $k_1$ and the
polarization for vector particles of ref.~\cite{hz} our results must agree,
in the limit of spinor masses going to zero, with those obtained with
 the method of ref.~\cite{hz} for every single diagram and every
polarization of external particles. This has been checked on several examples
and it may be used as a valid test of the correctness of the results.
\par
We believe that our method has some advantages with respect to those in
\cite{ks,hz}.
There is a certain similarity with the method of Hagiwara and Zeppenfeld
since both make use of $4 \times 4$ matrices, which however take indexes in
very different spaces. In our case only one $\tau$ matrix for each insertion
has to be computed, instead of one for each vertex and one for each propagator.
Moreover, we can freely choose the auxiliary vectors $k_0$ and $k_1$ which can
be useful both in simplifying the expressions generated in intermediate stages
and as a test. Compared with the method of ref.~\cite{ks,mana}, our formalism
is much more compact. It avoids the proliferation of terms generated by the
expansion of the momenta flowing in fermion propagators in terms of the
external momenta. Even with respect to the more efficient method suggested in
ref.~\cite{ks2} for treating internal propagators, we have a smaller number of
terms. The relationship between the matrix elements of u-tipe and v-type
spinors (\ref{Texpr},\ref{UM},\ref{Ti}) leads to simpler expressions than the
introduction of additional
auxiliary vectors, expecially when long fermion lines are present and the
insertions do not all correspond to external particles. As a consequence, it is
much easier in our formalism to keep track of partial results and to set up
recursive schemes of evaluation which compute and store for later use
subdiagrams of increasing size and complexity. Our method is also more flexible
in the choice of the polarization vectors for external vector particles and one
can avoid the extra integration needed to obtain the correct sum over
polarization of the formalism of ref.~\cite{ks,mana}. It allows to directly
compute cross sections with polarized W's and Z's for any desired polarization.
\par
The size of the matrix elements which have to be computed is steadily
increasing and, despite the fast improvements in computer performance, speed is
a vital issue of any method of doing calculations. We have compared our
formalism with those just mentioned, avoiding in all cases to use subroutine
and function statements which considerably increase computing time. Our
method was consistently two to four times faster.
\par
In conclusion, we have shown that it is possible to insert in spinor lines
completeness relations in order to diagonalize the operators $p\sla$ of
the propagators of momentum $p$. This is particularly convenient when one
uses the spinors \cite{ks}. We have presented the
formalism necessary to perform actual calculations, both in the massless and in
the massive case.

\newpage
\subsection*
{Appendix A}

\renewcommand{\theequation}{A.\arabic{equation}}
\setcounter{equation}{0}

Using the spinors (\ref{wspinor} \ref{uvks},\ref{uvksconj}),  their products
can be written as

\bea
\lefteqn{U(p_2,\lambda_2)\ol U(p_1,\lambda_1)= \frac{1}{4 \sqrt{2\,p_1\dotp
k_0}
\sqrt{2\, p_2\dotp k_0} }}\label{uubar}\\
& & \times(p\sla_2+M_2)\left[ (1+\lambda_1\lambda_2)-
(\lambda_1+\lambda_2)\gamma^5+k\sla_1[(\lambda_1-\lambda_2)-
(1-\lambda_1\lambda_2)\gamma^5]\right]k\sla_0(p\sla_1+M_1).\nonumber
\eea
Multiplying to the right by $\chi ^S$  or $\chi ^V$ of
eqs.~(\ref{chis},\ref{chiv}) , taking the trace  and with the help of
eq.~(\ref {Texpr}), one gets the following expressions for the functions
$A$, $B$, $C$, $D$ corresponding to a scalar and a vector insertion :

\bea\label{a2}
A^S_{+-}&=&c_l\left(k_0\dotp p_1\;k_1\dotp p_2-
          k_0\dotp p_2\;k_1\dotp p_1-i
\epsilon\,(k_0,k_1,p_1,p_2)\right)\nonumber\\
A^S_{-+}&=&c_r\left(-k_0\dotp p_1\;k_1\dotp p_2+
          k_0\dotp p_2\;k_1\dotp p_1-i
\epsilon\,(k_0,k_1,p_1,p_2)\right)\nonumber\\
B^S_{++}&=&c_r\; k_0\dotp p_2 \nonumber\\
B^S_{--}&=&c_l\; k_0\dotp p_2\nonumber\\
C^S_{++}&=&c_l\; k_0\dotp p_1\nonumber\\
C^S_{--}&=&c_r\; k_0\dotp p_1\nonumber\\
A^V_{++}&=&c_r\left(-k_0\dotp \eta\;p_1\dotp p_2+k_0\dotp p_1\;\eta\dotp p_2+
    k_0\dotp p_2\;\eta\dotp p_1+ i \epsilon\,(k_0,\eta,p_1,p_2)\right)\\
A^V_{--}&=&c_l\left(-k_0\dotp \eta\;p_1\dotp p_2+k_0\dotp p_1\;\eta\dotp p_2+
    k_0\dotp p_2\;\eta\dotp p_1- i
\epsilon\,(k_0,\eta,p_1,p_2)\right)\nonumber\\
B^V_{+-}&=&c_l\left(k_0\dotp \eta\; k_1\dotp p_2
            - k_0\dotp p_2 \; k_1\dotp \eta
         -i \epsilon\,(k_0,k_1,\eta,p_2)\right)\nonumber\\
B^V_{-+}&=&c_r\left(
        -k_0\dotp \eta\; k_1\dotp p_2 + k_0\dotp p_2 \; k_1\dotp \eta
        -i \epsilon\,(k_0,k_1,\eta,p_2)\right)\nonumber\\
C^V_{+-}&=&c_r\left(
        -k_0\dotp \eta\; k_1\dotp p_1 + k_0\dotp p_1\; k_1\dotp \eta
           +i \epsilon\,(k_0,k_1,\eta,p_1)\right)\nonumber\\
C^V_{-+}&=&c_l \left(
        k_0\dotp \eta\; k_1\dotp p_1 - k_0\dotp p_1\; k_1\dotp \eta
           +i \epsilon\,(k_0,k_1,\eta,p_1)\right)\nonumber\\
D^V_{++}&=&c_l\; k_0\dotp\eta\nonumber\\
D^V_{--}&=&c_r\; k_0\dotp\eta.\nonumber\\\nonumber
\eea
All functions $A$, $B$, $C$, $D$ for a single insertion not reported in the
preceding list are identically zero.
\par\noindent
The function $\epsilon$ is defined to be the determinant:
\be\label{a3}
\epsilon\,(p,q,r,s)= det \;
\left|\ba{cccc}       p^0 & q^0 & r^0 & s^0\\
                      p^1 & q^1 & r^1 & s^1\\
                      p^2 & q^2 & r^2 & s^2\\
                      p^3 & q^3 & r^3 & s^3 \ea\right|
\ee

\vfill\eject


\begin{thebibliography}{1}

\bibitem{jw} M.Jacob and J.C. Wick, \AP{7} (1959) 404.

\bibitem{bj} J.D. Bjorken and M.C. Chen,
\PR{154} (1966) 1335, O.~Reading-Henry \PR{154} (1967) 1534.

\bibitem{calkul}
P.~De~Causmaecker, R.~Gastmans, W.~Troosts and T.~T.~Wu,
\PL{B105} (1981) 215, \NP{B206} (1982) 53;
F.A.~Berends, R.~Kleiss,P.~De~Causmaecker, R.~Gastmans, W.~Troosts
and T.~T.~Wu,
\NP{B206} (1982) 61, {\bf B239} (1984) 382, {\bf B239} (1984) 395,
{\bf B264} (1986) 243, {\bf B264} (1986) 265;
For a review, see R.~Gastmans and T.~T.~Wu - The ubiquitous photon
(Addison Wesley, Reading, Ma,1990)


\bibitem{cr}
M. Caffo and E. Remiddi, \HPA{55} (1982) 339.

\bibitem{far}
G.R.~Farrar and F.~Neri, \PL{B130} (1983) 109.

\bibitem{gp}
G. Passarino, \PR{D28} (1983) 2867, \NP{B237} (1984) 249.

\bibitem{ks} F.A.~Berends, P.H.~Daverveldt and R.~Kleiss
\NP{B253} (1985) 441, R.~Kleiss and W.J.~Stirling,
\NP{B262} (1985) 235.

\bibitem{ks2} R.~Kleiss and W.J.~Stirling, \PL{B179} (1986) 159.

\bibitem{kg} J.~Gunion and Z. Kunszt,
\PL{B161} (1985) 333.

\bibitem{hz} K.~Hagiwara and D.~Zeppenfeld,
\NP{B274} (1986) 1.

\bibitem{cin}Z.~Xu, Da-Hua Zhang and L. Chang
\NP{B291} (1987) 392.

\bibitem{mana} C.~Mana and M.~Martinez,
\NP{B287} (1987) 601.

\bibitem{noi} A.~Ballestrero, E.~Maina and S.~Moretti,
{\it Turin Univ. preprint} DFTT 53/92, October 1992, to appear in
{\it Nucl.\ Phys.\ } {\bf B}.


\end{thebibliography}
\end{document}